\documentclass[11pt]{article}
	
	\newcommand{\blind}{0}
	
 \makeatletter
    \renewcommand\section{\@startsection{section}{1}{\z@}%
                                       {-3.5ex \@plus -1ex \@minus -.2ex}%
                                       {2.3ex \@plus.2ex}%
                                       {\normalfont\normalsize\fontfamily{phv}\fontsize{16}{19}\bfseries}}
    \renewcommand\subsection{\@startsection{subsection}{2}{\z@}%
                                         {-3.25ex\@plus -1ex \@minus -.2ex}%
                                         {1.5ex \@plus .2ex}%
                                         {\normalfont\fontfamily{phv}\fontsize{14}{17}\bfseries}}
    \renewcommand\subsubsection{\@startsection{subsubsection}{3}{\z@}%
                                        {-3.25ex\@plus -1ex \@minus -.2ex}%
                                         {1.5ex \@plus .2ex}%
                                         {\normalfont\normalsize\fontfamily{phv}\fontsize{12}{17}\selectfont}}
    \makeatother
	\usepackage{amsmath}
	\usepackage{graphicx}
    \usepackage{amsmath}
    \usepackage{amssymb}
	\usepackage{enumerate}
	\usepackage{xcolor}
    \usepackage{booktabs}
    \usepackage{tabularx}
    \usepackage{array}
    \usepackage[T1]{fontenc}
    \usepackage[utf8]{inputenc} 
    \usepackage{lmodern}
    \usepackage{textcomp}
    \usepackage{float}
    \usepackage{threeparttable} 
    \usepackage{indentfirst}
	\usepackage{natbib} 
	\usepackage{url} 
    \usepackage{algorithm}
    \usepackage[noend]{algpseudocode}
    \usepackage[super]{nth}
    \usepackage{hyperref}
    \usepackage{flafter}   
    \usepackage{placeins}
    \usepackage{needspace}   
    \usepackage{siunitx}
	\usepackage{todonotes}
    \usepackage[T1]{fontenc}
    \usepackage[left=0.7in,right=0.7in,top=0.75in,bottom=0.75in]{geometry}

    \usepackage[utf8]{inputenc}
	\usepackage{amsfonts, amsthm, latexsym, amssymb}
	\usepackage{lipsum} 
        \usepackage{tablefootnote}
	\newcolumntype{L}[1]{>{\raggedright\arraybackslash}p{#1}}
    \newcolumntype{C}[1]{>{\centering\arraybackslash}p{#1}}
\begin{document}

		\def\spacingset#1{\renewcommand{\baselinestretch}%
			{#1}\small\normalsize} \spacingset{1}
        \if0\blind
		{
			\title{\bf Leveraging High-Fidelity Digital Models and Reinforcement Learning for Mission Engineering: A Case Study of Aerial Firefighting Under Perfect Information
   }
			\author{İbrahim Oğuz Çetinkaya$^1$, Sajad Khodadadian$^1$ and Taylan G. Topcu$^1$ \\
            $^1$\small Grado Department of Industrial and Systems Engineering, Virginia Tech, Blacksburg, VA }
			\maketitle
	     }  \fi

		\if1\blind
		{

            \title{\bf \emph{IISE Transactions} \LaTeX \ Template}
			\author{Author information is purposely removed for double-blind review}
			
\bigskip
			\bigskip
			\bigskip
			\begin{center}
				{\LARGE\bf \emph{IISE Transactions} \LaTeX \ Template}
			\end{center}
			\medskip
		} \fi
		\bigskip

	\spacingset{1.5} 
\vspace{-1cm}

\begin{abstract}
As systems engineering (SE) objectives evolve from design and operation of monolithic systems to complex System of Systems (SoS), the discipline of Mission Engineering (ME) has emerged which is increasingly being accepted as a new line of thinking for the SE community. Moreover, mission environments are uncertain, dynamic, and mission outcomes are a direct function of how the mission assets will interact with this environment. This proves static architectures brittle and calls for analytically rigorous approaches for ME. To that end, this paper proposes an intelligent mission coordination methodology that integrates digital mission models with Reinforcement Learning (RL), that specifically addresses the need for adaptive task allocation and reconfiguration. More specifically, we are leveraging a Digital Engineering (DE) based infrastructure that is composed of a high-fidelity digital mission model and agent-based simulation; and then we formulate the mission tactics management problem as a Markov Decision Process (MDP), and employ an RL agent trained via Proximal Policy Optimization. By leveraging the simulation as a sandbox, we map the system states to actions, refining the policy based on realized mission outcomes. The utility of the RL-based intelligent mission coordinator is demonstrated through an aerial firefighting case study. Our findings indicate that the RL-based intelligent mission coordinator not only surpasses baseline performance but also significantly reduces the variability in mission performance. Thus, this study serves as a proof of concept demonstrating that DE-enabled mission simulations combined with advanced analytical tools offer a mission-agnostic framework for improving ME practice; which can be extended to more complicated fleet design and selection problems in the future from a mission-first perspective.
\end{abstract}

\section{Introduction} 
\label{s:intro}
The discipline of systems engineering (SE) is concerned with the successful realization and lifecycle management of complex engineered systems \cite{Kossiakoff2011} by balancing competing performance and functional characteristics within the boundary of a given engineered system (e.g., an aircraft, missile, automobile) \cite{Walden2023}. However, modern engineering challenges rarely exist in isolation and in many cases, they operate within large scale, collaborative networks of systems also denoted as System of Systems (SoS). Unlike monolithic engineered systems, SoS are characterized by the operational and managerial independence of their constituent systems, evolutionary development, geographical distribution, and complex emergent behaviors arising from a rapidly growing web of interactions \cite{gorod2008, maier1998}. \\
\indent Extending SE concepts to SoS necessitate some adaptations \cite{jamshidi2008} given the need to shift the objective from optimizing monolithic systems to orchestrating interactions between a heterogenous set of constituent systems to achieve capabilities beyond their individual scopes \cite{keating6, keating7, nielsen2015}. Prior work in this area has made substantial progress on designing better constituent systems and architectures for various application areas such as aerospace \cite{oneill2010, selva2017, grogan2016}, disaster management \cite{fan2020}, supply chains \cite{choi2018, jaradat2017} and among many others \cite{delaurentis2005, raz2020, mogahed2025}. In many SoS contexts, diverse sets of assets must be selected, configured, and coordinated to serve an evolving set of missions. While SoS architecting and evaluation methods are relatively more mature, deployed SoS architectures remain brittle as they are costly to change once implemented and their performance is susceptible to operational and contextual changes \cite{raz_conceptual2024, sapol2022}. Consequently, ensuring effective SoS operations under future operational uncertainties remains a research challenge across domains \cite{raz2020, garvey2014, cherfa2019}. There is a pressing need for analytically rigorous methods to analyze SoS capabilities \cite{mcmanus2005}. \\
\indent To this end, architecting an SoS consisting of heterogeneous consistent systems that may dynamically evolve in terms of their roles and integration level depending on mission conditions \cite{mitre2024}. ME is defined as the “deliberate planning, analyzing, organizing and integration of current and emerging operational and system capabilities to achieve desired mission outcomes” \cite{hutchison_mission_2018}. Two tightly coupled decision problems emerge in ME. First, is the selection of assets, by addressing queries such as which assets to deploy to the field, with what configuration, and payload. The second is the management of the selected assets during the course of the mission, looking into how to command, control, and dynamically re-task assets under various mission conditions. \\
\indent Consequently, the inherent analytical challenge of ME is formidable particularly given that mission environments are uncertain, often only partially observable, dynamic; and also, the mission constraints are of combinatorial nature \cite{keating7, nielsen2015}. Raz et al. summarize these ME challenges in three pain points: (i) analytic methods to cope with the vast SoS design spaces; (ii) adaptive task allocation and reconfiguration of the SoS in response to evolving contexts; and (iii) principled identification of multiple configurations achieving a good balance rather than single-point optima \cite{raz_conceptual2024}. In addition to these concerns, mission asset interactions generate nonlinear relationships and influence mission performance; which in return render the use of traditional simulation based analytical tools such as tradespace exploration methods less effective. This is because even when tractable mission models exist which can be solved in reasonable times, solving them with classical numerical methods at an operational timeline is computationally burdensome. Finally, the vast solution space renders exhaustive approaches infeasible \cite{dachowicz_mission_2021, hazelrigg_framework_1998}. \\
\indent These challenges call for adaptive, data-driven approaches that can learn integrated ME policies for both asset selection and management. Previous studies utilize uncertainty quantification, various machine learning techniques \cite{raz2020, dachowicz_mission_2021}, interval algebra \cite{raz_conceptual2024}, graph theory \cite{covello2023} and, optimization combined with digital twins \cite{dtRoute}. However, operationalizing such methodologies necessitates rich data on missions, constituent systems, and their interdependencies. According to NASA, ME includes the problem of integrating development activities across engineering and operations \cite{poza_mission_2015}. To work towards solving this problem, Digital Engineering (DE) is proposed as an approach for establishing a consensus between data-driven infrastructures, methodologies, innovation and workforce \cite{henderson2023} to deliver high pay-off solutions in a convenient timeline. Concurrently, DE practices such as model-based systems engineering, high-fidelity simulations, data pipelines from IoT/telemetry, and digital twins \cite{dtRoute} can serve as the infrastructure that can enable addressing core ME needs with analytical tools \cite{d_of_defense_mission_2020}. A high fidelity DE infrastructure could serve as a testbed for testing ME policies against realistic scenarios before they are deployed to the field. \\
\indent This paper addresses this challenge at its core, by proposing an analytical methodology that integrates high-fidelity digital mission models with reinforcement learning (RL) techniques for intelligent mission coordination. More specifically, we address the following research question: how can RL methods be synergistically integrated with high-fidelity mission models for ME to enable intelligent and robust mission coordination given mission uncertainties? Recent work in this area pursued generation of AI-driven tactics and their explainability for ME \cite{dachowicz_mission_2021}, resource orchestration of multi-agent coordination \cite{chen2024, chen2026}, and arrangement of rewards for adaptive tasks \cite{ji2025}. Hence, this work adds to the growing body of literature that aims to leverage artificial intelligence for SoS. Additionally, our work targets the second item in Raz’s list: adaptive allocation/reconfiguration within collaborative or acknowledged SoS, where independence of constituent systems and changing operational conditions make static designs underperform \cite{raz_conceptual2024}. \\
\indent We address the posed research question by formulating the mission execution as a Markov Decision Process and then applying Proximal Policy Optimization (PPO), which is an RL-based algorithm \cite{schulman2017}. By interacting with the simulation environment over thousands of iterations, the intelligent mission coordinator maps system states to actions, guided by reward signals reflecting the achievement of mission objectives. We then illustrate the utility of our approach on a notional aerial firefighting mission that was developed by DLR \cite{cigal2025}. Here, task is to mitigate the wildfire in a given region by deploying a set of aerial firefighting assets, and managing their tactics under the assumption of perfect and real-time information. \\
\indent Our findings are two-fold. First, once trained, the intelligent mission coordinator significantly outperforms the baseline mission performance regardless of mission uncertainties. Second, the mission coordinator also significantly reduces the variability of mission performance. These findings suggest that integrating mission models with RL could balance the computational burden of ME and could help serve as an analytically rigorous approach for ME. While this paper utilizes an aerial firefighting case study, the proposed concepts are mission-agnostic and adaptable to other operational settings. Ultimately, we demonstrate that a DE-enabled mission simulation, combined with advanced analysis, offers a significant advancement to SE and ME practice.

\section{Methodology}
\begin{figure}
\centering
\includegraphics[width=1.0\linewidth]{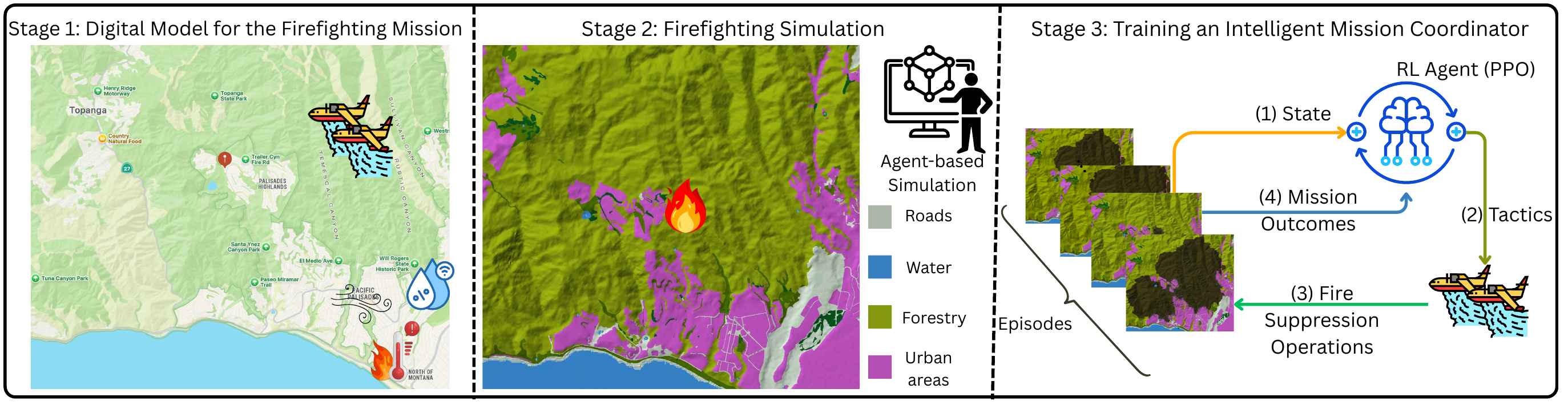}
\caption{Methodological Overview}
\label{fig:overview}
\end{figure}

In this study, we propose a method that integrates high-fidelity digital mission representations with RL for intelligent mission coordination. More specifically, the proposed approach aims to make real-time tactical decisions based on the status of the mission given a fixed set of mission assets. We demonstrate the value of the proposed method by implementing it on a notional aerial firefighting mission, utilizing a mission model that was developed by the German Aerospace Center (DLR) \cite{cigal2025}. The authors express their gratitude for the DLR researchers, namely Dr. Prajwal Shivaprakasha, Mr. Nabih Naeem and Mr. Nikolaos Kalliatakis who have facilitated the mission model and simulation as part of the Colossus Challenge. Figure \ref{fig:overview} provides an overview of our approach that is composed of three stages. The first stage is formulation of a digital mission model, that includes physics-based representations of the mission’s geographical area, along with the threat, mission assets, laws of physics, and relevant elements of the environment (e.g., vegetation type and density, wind, temperature, humidity). Second step is the mission simulation using this model. In our case the mission simulation is based off of DLR’s agent-based simulation that helps simulate the interaction of the mission assets with the given threat and the environment. We leverage this simulation as a testbed for measuring the outcomes of various tactical decisions. The third step is the design, implementation, and training of a RL-based centralized mission coordination agent. This RL agent experiments with tactics for a fixed set of mission assets, and learns from mission outcomes to infer more preferable tactics over time. Training enables the RL agent to react to new missions with agility, and select a new set of tactics that significantly outperform baseline tactics. Below, we discuss each of these steps in detail. 

\subsection{Firefighting Mission Model}
\label{s:mission_model}
Our study utilizes a digital mission model and an accompanying agent-based simulation to enable intelligent and integrated tactical decision making under uncertain mission conditions for aerial firefighting missions. The model and the accompanying agent-based simulation is developed to analyze missions and inform the design of SoS to realize mission outcomes in a firefighting situation. \\
\indent The model represents a geographical area by using a grid structure; each cell is characterized with a set of states. A cell can be non-flammable if it does not have any fuel, these cells are represented using gray (i.e., roads) and blue (i.e., water sources) in the map. On the other hand, a cell can be flammable such as urban areas (purple) and forestry (green). Forestry cells are further detailed to reflect varying combustibility; they incorporate specific fuel types -such as pines, leaf litter, needles and fallen leaves- to accurately model how fire propagates through different vegetation. Flammable cells can be present in one of the following states that reflect different phases of burning: combustible, early burning, full burning, extinguishing, and burnt. Airports are also represented in the digital model to serve as bases for firefighting operations. \\
\indent Moreover, the digital model constitutes a basis for an agent-based simulation that replicates a wildfire breaking out in the targeted region. The dynamic operational context in the simulation includes changing wind direction and speed, temperature, and humidity over time; all of which significantly impact wildfire propagation. Temperature, and humidity are defined as ranges. To replicate the weather fluctuations of a single day, the simulation interpolates these values over time, peaking in the middle of the day. Wind speed is a scalar input, and it is fluctuated using a sine function to simulate natural variability. Finally, the wind degree follows a random uniform distribution. Together, wind speed and degree constitute the main uncertainty in fire propagation as various angle and speed combinations of wind result in significantly divergent behaviour. In this firefighting simulation, a pre-determined fleet is tasked to suppress the fire with assigned tactics for each aircraft. \\
\indent At the end of each simulation four main metrics are reported to evaluate mission outcomes: (i) Casualties (number of lost lives), (ii) Cost of Burnt Area (€), (iii) Burnt Area (m2), and (iv) Emissions of Burnt Area (tonnes CO2). These outcome metrics are then aggregated into a value function to measure the success of the mission. These distinct combustible cells depicted in Figure \ref{fig:overview} are not merely visual; they contribute differently to the simulation's performance metrics. For example, while urban areas are heavily weighted in terms of casualties and costs, burning forests are the primary drivers of emission metrics.

\subsection{Markov Decision Process Formulation and RL Implementation}
Markov Decision Processes (MDP) provide mathematical representations for decision making under uncertainty, specifically capturing dynamic scenarios where an agent's actions determine the next state of the environment. An MDP representation consists of an environment which can be present in a set of states $(s)$ that reflect different environmental conditions, and an agent that can interact with the environment through its actions $(a)$. The environment transitions from one state to another as the agent manipulates it through its actions. Transitions between states occur stochastically and are expressed as transition probabilities from $s$ to $s'$ given that action $a$ is undertaken, $P(s' \mid s, a)$. In each state, the agent receives a reward signal associated with the current transition, $R(s,a,s')$, and it strives to maximize the expected long-term reward, discounted over time. To achieve this goal, the agent trains a policy function $\pi(a \mid s)$, which maps states to decisions over time. Finally, the policy is improved by ascending the gradient of the expected value of the discounted return. \\
\indent Deep Reinforcement Learning methods have developed significantly since the introduction of Deep Q-Networks \cite{mnih2015}, which utilize value-based learning to learn high-dimensional discrete control tasks. However, value-based methods are documented to struggle with convergence in stochastic environments compared to policy-gradient methodologies \cite{sutton1999}. To improve stability, the Trust Region Policy Optimization (TRPO) approach introduced a constraint to guarantee monotonic policy improvement, relying on computationally expensive second-order optimization \cite{schulman2015}. Moreover, this study utilizes Proximal Policy Optimization (PPO) \cite{schulman2017} to benefit from its stability and ease of parameter tuning compared to other methods. This method is particularly well suited for our case as it caters to the need to learn complex stochastic policies for tactical decision making without overlearning. \\
\indent PPO is a policy-gradient algorithm that enhances stabilization (i.e., prevents poor performance caused by excessively large updates) during learning by constraining each policy update to be close (i.e., proximal) to the previous policy \citep{schulman2017}. Therefore, sudden jumps in the policy and hence overlearning are avoided, providing a more stable learning behaviour overall. In this way, the exploration--exploitation phase of the learning process is kept under control. To achieve this, PPO considers the ratio (\ref{eq:eq_1}) between the updated policy $\pi_{\theta}(a_t \mid s_t)$ and the current policy $\pi_{\theta_{\mathrm{old}}}(a_t \mid s_t)$ to maximize the clipped surrogate objective (\ref{eq:eq_2}). Maximizing the clipped surrogate objective in equation~(2), avoids excessively large policy updates and ensures that the policy is trained steadily in a controlled manner where the `clip' operation forces the change to be within the range $[1-\varepsilon,\, 1+\varepsilon]$, where $\varepsilon$ is a small number.
\begin{equation}
r_t(\theta) = \frac{\pi_{\theta}(a_t \mid s_t)}{\pi_{\theta_{\mathrm{old}}}(a_t \mid s_t)}
\label{eq:eq_1}
\end{equation}

\begin{equation}
L^{\mathrm{CLIP}}(\theta)
= \mathbb{E}_{t}\!\left[
\min\!\Big(
r_t(\theta)\,\hat{A}_t,\;
\mathrm{clip}\!\big(r_t(\theta),\,1-\varepsilon,\,1+\varepsilon\big)\,\hat{A}_t
\Big)
\right]
\label{eq:eq_2}
\end{equation}

\subsection{Implementation of the RL Agent}
The implementation of the centralized mission coordination RL agent for the given mission problem is as follows. The agent is responsible for managing the operations of a given fleet on the tactical level under perfect information. At each pre-determined timestep during the simulation, the RL agent observes the state (in terms of the operational context and the assets’ states) and determines an action with its policy. After the operation is completed and the simulation proceeds to next timestep, agent recognizes the feedback by the change in mission outcomes; the change in Measure of Effectiveness that maps the overall damage inflicted throughout the fire to an interval in $[-1, 1]$. Below, we discuss the four main components of the RL agent, the state space, the action space, the reward function, and its training via simulation. 

\subsubsection{Mission Scenario Analyzed in this Paper}
As a representative case study, this study adopts the aforementioned mission model for the Palisades wildfire of January $7^{\text{th}}$, 2025. This catastrophic event devastated the coastal California town, burning 95 km$^2$, causing over 25 billion USD of damage, and resulting in 12 deaths over a 31-day period \cite{babrauskas_palisades_2025}. The region's geography, characterized by ocean proximity, inland lakes suitable for water scooping and heterogeneous structure composed of urban area and forestry presents a complex mission environment. We analyze a scenario in which a wildfire breaks out near the north of town and detected after an hour with the simulation calibrated to replicate the temperature, humidity and wind speed of the event. After the detection, a firefighting fleet consisting of two large air tankers (DHC 515) is deployed to the area to supress the fire within 16 hours which is the average time of the fire to reach boundaries of the given region. The aircraft takes off from airports outside the town, repeatedly load water from available sources, conduct suppressing operations and refuel as needed. For each move, aircraft decide their next actions based on the tactics they are assigned. Here, it is important to note that we purposefully limited the fleet to just two tankers as this paper aims to propose a proof of concept for RL-based mission coordinator for a limited fleet before we extend these ideas to more realistic representation of heterogenous mission assets. We recognize that the actual firefighting force was drastically different in the actual Palisades disaster. 
\subsubsection{State Space}

States of the mission simulation reflects the status of the mission and serves as a means for the RL agent to analyze the mission environment and the current state of the fleet. Hence, the state space is composed of 16 environment variables and 6 variables per asset to describe the status of the mission assets in each time step which are provided in Table \ref{tab:statevars}. 

\begin{table}[H]
\centering
\small
\caption{Description of State Variables}
\label{tab:statevars}
\begin{tabularx}{\linewidth}{|p{0.28\linewidth}|X|p{0.22\linewidth}|}
\hline
\textbf{Variable} & \textbf{Description} & \textbf{Unit of Measurement}\\
\hline
Ambient temperature & Current environmental temperature & Celsius degrees\\ \hline
Relative humidity & Current environmental humidity & Percentage\\ \hline
Wind speed and degree & Current wind speed and degree & m/s and degrees\\ \hline
Fire's distance to water and the fire line & Shortest distance from current fire front to the indirect fire line segment being built now and to the closest water resources & Meters\\ \hline
Active fire front count & Number of currently burning cells & Cells\\ \hline
Burnt area fraction & Burnt area divided by total area & Dimensionless\\ \hline
Time to sunset & Duration until sunset & Minutes\\ \hline
Fire center & Centroid of all burning cells in the map & X, Y coordinates\\ \hline
Fire spread angle & Propagation of the fastest-spreading cell relative to the centroid & Degrees\\ \hline
Fire distance to boundaries & Distance between four edges of the region and the fire center & Meters\\ \hline
Location of the asset & Current location of the aircrafts & X, Y, Z coordinates\\ \hline
Payload binary flag & Does the agent carry fire suppressor & Dimensionless\\ \hline
Propellant fraction & Remaining propellant left in the fuel tank of aircraft & Dimensionless\\ \hline
Return margin & Remaining propellant fraction excluding the mandatory reserve required to reach the nearest airbase & Dimensionless\\ \hline
\end{tabularx}
\end{table}

\subsubsection{Action Space}
The mission coordination agent observes of the mission through the state variables and specifies the tactical orders at each timestep. Hence the action space of the RL agent is composed of tactics (i.e., instructions on how to suppress the fire) for each mission asset. Tactics in this case consist of triplets that specify:
 \begin{itemize}
        \item \textbf{Select Point of Interest (POI)} An aircraft can be ordered to prioritize protecting various strategic locations and determine POIs accordingly. These POIs are: water resources, areas with highly combustible vegetation, areas where upslope spread is likely, and determining POI to indirectly intervene, containing the fire in an ellipse.
        \item \textbf{Track POI} This element of the triplet determines the method to track the selected POI. There are three alternatives: direct tracking, indirect tracking, and following the fire front. Using direct tracking, aircraft does not change its course under any condition and follows the selected POI. In indirect tracking, aircraft checks if the POI is burnt or suppressed, and if it is, aircraft recalculates a new POI. Finally, “follow the fire front” tracking means the aircraft will check the spread of fire around the POI, and target the point with the highest spread rate. With this option, it is possible to contain the fire cells with highest spread rate; however, certain amount of the fire suppressant is dropped to areas that are not actively burning. 
        \item \textbf{Suppress Mode} This element of the triplet defines the method of suppressing the fire. There are two suppression options: direct or indirect. With direct suppressing, the aircraft approaches the POI by orienting its suppression angle to maximize burnt area captured while minimizing already suppressed areas. On the other hand, with indirect suppressing, the agent orients its suppression angle to prioritize the fulfilment of the fire line by dropping the suppressant in an tangential angle to the spread of the fire. 
    \end{itemize}

\begin{figure}
\centering
\includegraphics[width=1.0\linewidth]{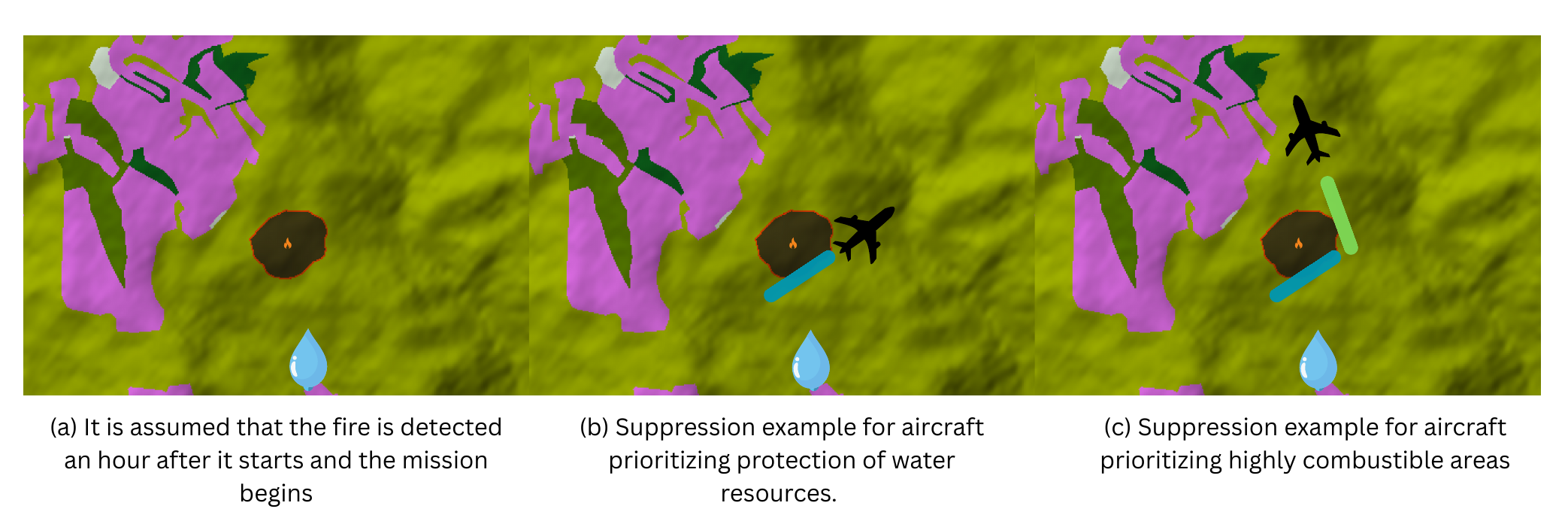}
\caption{Simulation progression, and two example track and suppress operations}
\label{fig:tactics}
\end{figure}

\indent These triplets result in 24 tactics (4 Select POI x 3 Track POI x 2 Suppress) for each aircraft. Simulation takes the tactic triplets as input for each timestep and it continues operations until either (i) the fire is contained or (ii) the time limit is reached which varies from region to region or (iii) the fire could not be contained and reached town borders, meaning it will spread to nearby towns and will require further intervention, possibly with larger fleets. \\
\indent To visualize how these POI decisions on the mission simulation, we present Figure \ref{fig:tactics}. Figure \ref{fig:tactics}.a illustrates the moment that the fire is first detected; the green area shows the vegetation, the black represents the area that is already burnt, the red shows the line of fire, and the blue water drop illustrates the relative location of water resources on the map. Figures \ref{fig:tactics}.b and \ref{fig:tactics}.c shows two Track POI options, “prioritize water resources” and “prioritize highly combustible vegetation areas” with direct suppression. These track POI options are indicated with blue and green suppression actions shown in lines, respectively. Figure \ref{fig:tactics} also allows us to illustrate the complexity and challenge of intelligent mission coordination. As each aircraft takes actions according to the tactic they are assigned, the propagation of the fire is affected by these tactical decisions as much as they are by the conditions of the environment at the time such as wind, temperature, and humidity; and the type of cell the fire is spreading to (e.g., highly flammable vegetation vs. urban areas). Collectively, these influence the goodness of mission outcomes as represented by the Measure of Effectiveness which is discussed next. Thus, mission engineering can significantly benefit from rigorous analytical methods and digital testbeds to analyze mission uncertainties and develop the ability to rapidly generate novel tactics given the state of the fleet and mission conditions at the time.

\subsubsection{Reward}
Our methodology utilizes a reward structure based on a notional Measure of Effectiveness (MoE), that serves as a value function to capture the damage inflicted by the wildfire. The notional MoE for this mission is calculated as follows: 

\begin{equation}
\mathrm{MoE}
= 1 - 0.25\left(
\frac{BA_i}{MBA}
+ \frac{CA_i}{MCA}
+ \frac{E_i}{ME}
+ \frac{C_i}{MC}
\right) - \mathrm{outofBound}
\label{eq:moe}
\end{equation}

\indent $BA_i, CA_i, E_i$ and, $C_i$ are burnt area throughout the fire, cost of the area burnt, emissions resulting from the burnt area and finally the casualties for period i. Furthermore, $MBA, MCA, ME$ and, $MC$ are the maximum possible damage that can be inflicted given the specific scenario, bounded by the region’s characteristics (e.g., max casualties are equal to the population of the Palisades region). This allows the MoE to scale the damage to the environment in terms of a combination of factors in the range [-1,1]. Finally, $outofBound$ is a binary variable that equals 1 if the fire spreads to an adjacent region and 0 if it stays within the region after 16 hours. Aggregating the metrics, MoE provides a means to evaluate the mission’s effectiveness; values close to 1 indicate successful missions, while values closer to -1 indicate poor mission performance.
\subsubsection{Simulation Workflow for Training the RL Agent}
\begin{figure}[H]
\centering
\includegraphics[width=1.0\linewidth]{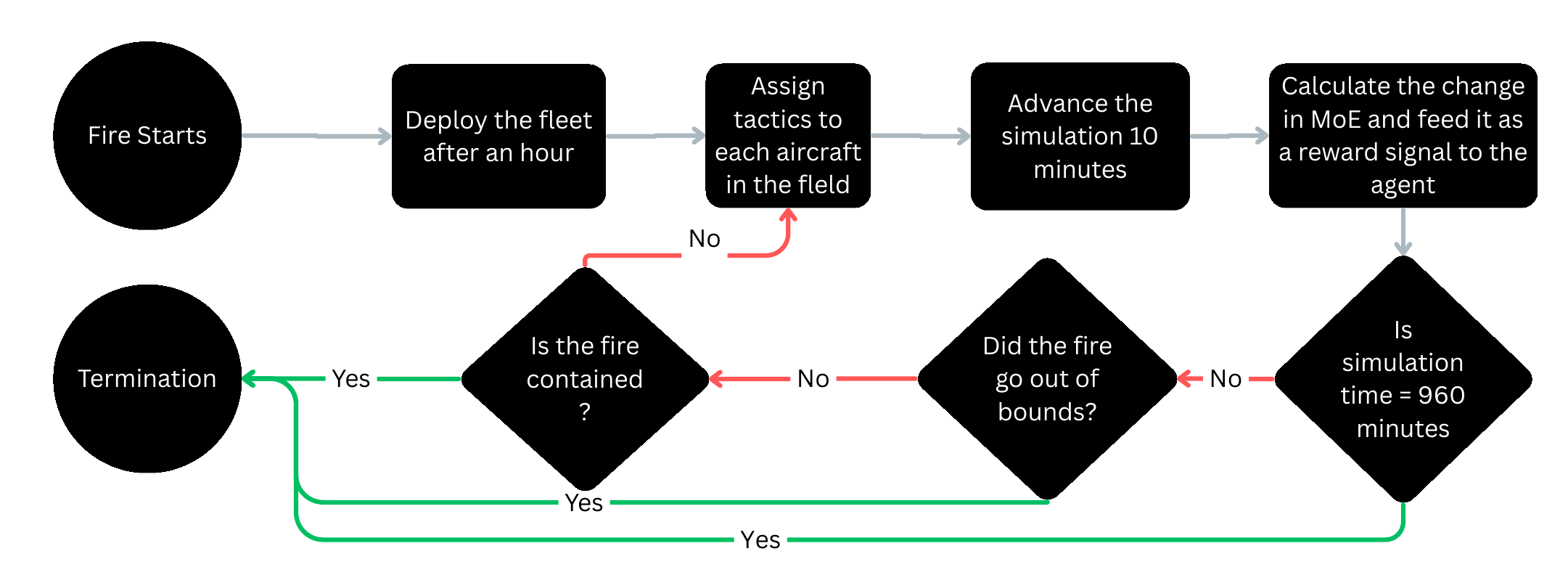}
\caption{Operational workflow of the simulation environment for training the RL-based intelligent mission coordination agent }
\label{fig:op_workflow}
\end{figure}

Now that we have all the pieces of the RL agent, we discuss the implementation and training of the RL agent via the mission simulation. The simulation advances in discrete 10-minute time steps which balance the data collection and training time and also are consistent with the inherent nature of firefighting operations. Every 10 minutes, the agent assesses the state of mission by using the state variables, and decides on the tactics assigned to each mission asset (i.e., aircraft) for that timestep in the simulation. Then, the simulation proceeds with the assigned tactics and records the changes in its environment. The simulation tallies the environmental damage by using the change in the MoE, and uses this as its reward signal. In other words, at each timestep, the RL agent receives the delta change as its reward $r_t = \Delta \mathrm{MoE}_t = \mathrm{MoE}_t - \mathrm{MoE}_{t-1}$. After the RL agent receives the reward signal, a cycle is completed and the next one starts. This cycle is repeated until termination of the simulation which is limited at 16 hours (960 minutes = 96 decision intervals). As a result, PPO learns a policy that maximizes expected discounted sum of $\Delta \mathrm{MoE}_t$ over episodes which is expressed as the following value function: $V^{\pi}(s_0) = \mathbb{E}_{\pi}\!\left[\sum_{t=1}^{T-1} \gamma^{t-1} \,\Delta \mathrm{MoE}_t \mid s_0\right]$. \\
\indent To enhance stability, the PPO algorithm employs mini-batch gradient descent, a technique widely recognized for reducing the variance of error estimates \citep{schulman2017}. In our implementation, each complete 96-step episode functions as a single batch because of the 16-hour limit to the simulation, which is further subdivided into the mini-batches of 32 steps. The policy parameters are updated incrementally using these smaller subsets; this approach ensures the agent develops robust strategies while minimizing the risk of overfitting to specific simulation instances. \\
\indent Figure \ref{fig:op_workflow} displays the training process over a single fire simulation (an episode); circles represent the initiation and termination of the simulation, boxes are used for operations during the simulation, and diamonds are used to indicate decisions to terminate or not. The fleet is deployed an hour after ignition and the RL agent assigns tactics to aircraft, then the change in MoE is calculated and transmitted back to the agent as feedback. If the simulation is not terminated for the reasons listed in Section \ref{s:mission_model}, the cycle repeats: agent makes new tactical decisions for aircraft and this process continues until simulation termination.

\section{Results}
\begin{figure}[H]
\centering
\includegraphics[width=1.0\linewidth]{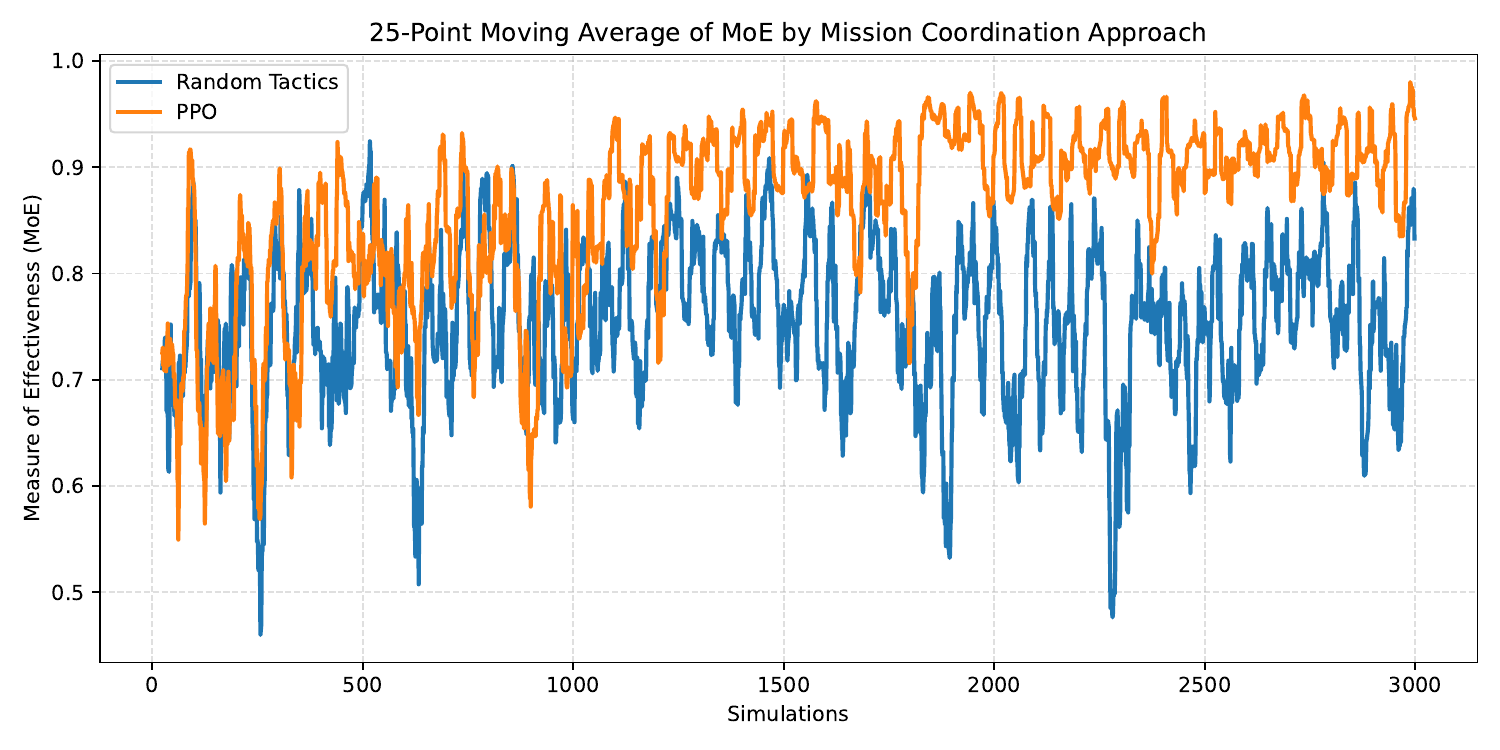}
\caption{Operational workflow of the simulation environment for training the RL-based intelligent mission coordination agent }
\label{fig:ma_3000}
\end{figure}

In this section, we present our findings of the RL based intelligent mission coordinator’s performance which assigned tactics to each mission asset at every 10-minute time interval. We start our presentation with Figure \ref{fig:ma_3000}, that portrays the moving average of MoEs over the training period of 3,000 simulations (i.e., episodes). While we acknowledge that extended training is often required to fully exploit RL capabilities, we restrict this study to 3,000 episodes to establish a proof of concept, balancing performance and the high computational cost of training. In Figure \ref{fig:ma_3000}, the x-axis represents the epochs, and the y-axis represents the MoE, where orange stands for the RL approach and blue represents the Random Tactics assigned to aircraft in each time step that serves as a benchmark for the RL agent’s performance. A moving average with a window size of 25 simulations was applied to smooth the noise in the training data. 

\begin{figure}[H]
\centering
\includegraphics[width=1.0\linewidth]{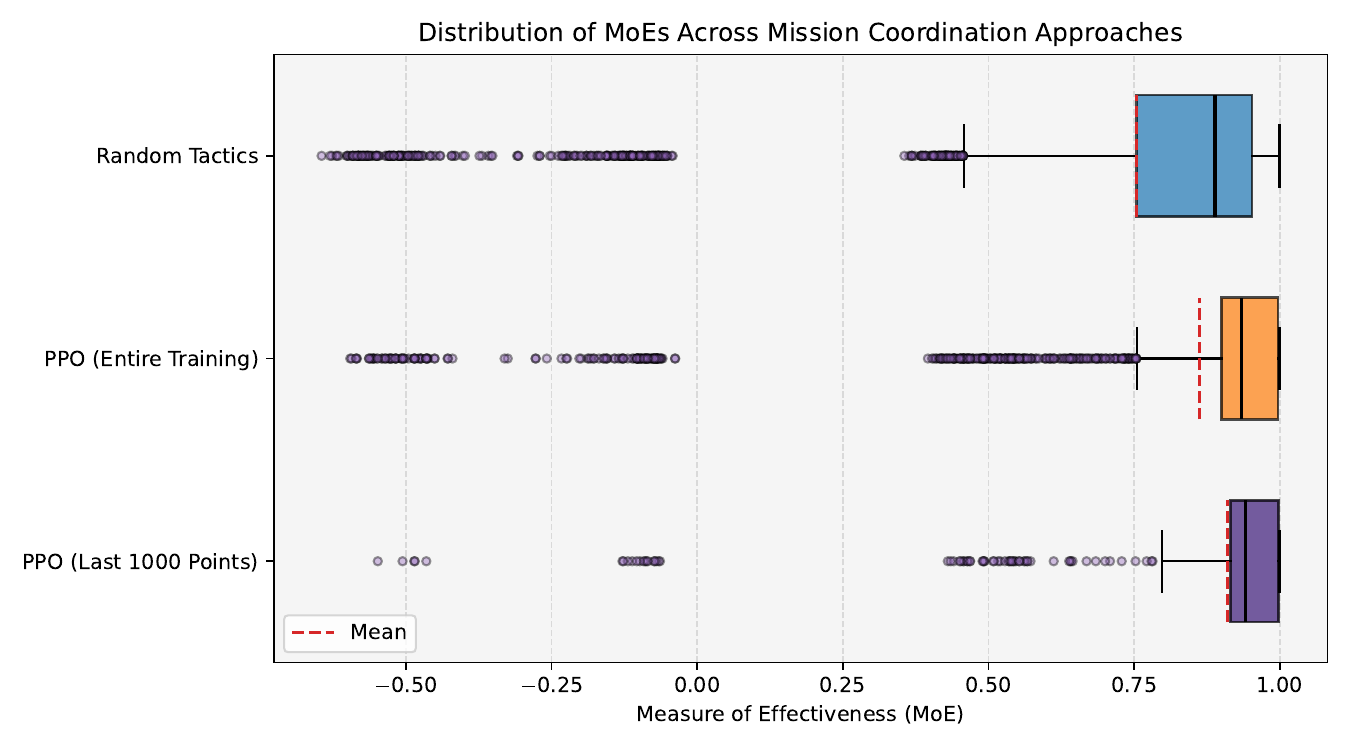}
\caption{Operational workflow of the simulation environment for training the RL-based intelligent mission coordination agent }
\label{fig:boxplots}
\end{figure}

\indent A review of Figure \ref{fig:ma_3000} suggests that while both approaches exhibit oscillating behaviour initially, PPO begins to outperform “Random Tactics” around 1,200th iteration, and retains this performance gap until termination. Beyond this step, PPO stabilizes around an MoE of 0.91 with significantly lower oscillation compared to the baseline. Although minor fluctuations remain near termination, these are not due to learning instability, but rather the inherent stochasticity of the simulation. Given the probabilistic nature of the simulation where factors like fire severity are inherently variable, even a fully optimized policy will experience fluctuations. \\
\indent To further analyze the learning behavior, we generated a box-plot that focuses on the RL agent’s performance towards later stages of training. Figure \ref{fig:boxplots} demonstrates the distribution of MoE values for Random Tactics (blue), PPO for the entire training period (orange), and finally the stabilized PPO agent during the last 1,000 simulations (purple). Dotted red lines mark the mean for each approach. Figure \ref{fig:boxplots} highlights a distinct performance gap: PPO achieves a higher median compared to random tactics, and its performance is considerably better in the final 1,000 episodes of training. Furthermore, PPO demonstrates superior stability. While the random tactics exhibit with frequent outliers, PPO successfully minimizes these anomalies over time, with the stabilized agent points portraying a significantly tighter distribution than the full training set. Perhaps more importantly, these performance gaps are statistically significant. Random Tactics achieve a median MoE of 0.888, while PPO achieves a significantly higher median of 0.935. Although the Random Tactics is surpassed by PPO, it exhibits somewhat high MoE. We attribute this to having two large air-tankers that provide high fire-suppression capability that provide a high ceiling.

A statistical comparison of the RL-based mission coordination agent against the Random Tactics using the Mann--Whitney U test confirms the difference is statistically significant, where the PPO agent achieves a mean MoE of 0.862 compared to 0.755 with a $p$-value of $<0.001$. Remarkably, the PPO distribution displays a concentration of high scores in the upper quartiles, indicating better performance. Outliers at the left tail of the PPO distribution are due to the exploration effort of the RL agent, which are inherent to the training process. RL agent deliberately opts for suboptimal solutions to explore different policy alternatives and avoid being stuck in local optima. The second point we would like to highlight is that, PPO is significantly tighter than the Random Tactics performance after it stabilizes as shown in the purple box plot in the final 1{,}000 episodes. Mann--Whitney U test confirms that the stabilized RL agent yields a significantly higher MoE, with a mean of 0.911, compared to 0.755 for the Random Tactics episodes ($U \approx 7.03 \times 10^{5}$, $p < 0.001$). Furthermore, the training process demonstrates notable improvement over time, as the final 1{,}000 PPO episodes outperforms the first 1{,}000 episodes (mean MoE: 0.911 vs.\ 0.784; $U \approx 6.55 \times 10^{5}$, $p < 0.001$). Finally, we observe that inter-quartile ranges are getting narrower as we train PPO more and both datasets have lower spread than the baseline. These results indicate that the PPO agent not only learns better tactics compared to the baseline but also continues to refine its policy throughout the training.

\section{Discussion}

\indent As documented by numerous US Government Accountability Office Reports, SE community has notoriously struggled with dealing with local problems that are constrained within the boundary of a standalone system \cite{schwenn2011_44, oakley2018_45, us2019_46, oakley2017columbia_47}. Methods and tools for dealing with ME complexity, albeit a few exceptions \cite{chen2024, chen2026}, is nascent. To that end, this paper proposed to integrate high-fidelity digital mission models with RL algorithms for intelligent mission coordination given operational uncertainties. This paper provided the proof of concept for being able to do this effectively, for fixed set of mission assets, and under the restrictive assumptions of perfect information. Our finding suggest that this is a useful inquiry, as the RL-based intelligent mission coordinator significantly outperforms Random Tactics and leads to increased stability in both mission performance and elimination of lower-bound catastrophic failures despite deviations in operational conditions. The reduction in performance variability is critical for mission-critical systems, where predictability is as valuable as raw performance. This finding also it confirms that the observed improvements are the result of learning rather than sporadic good performances driven by the simulation’s randomness. We also found that this performance gap widens as the agent is trained longer and the clear upward trend in performance indicates and ongoing optimization phase suggesting that longer experiments would likely yield positive returns in agent performance. \\
\indent While the results validate the potential value of the proposed methodology, several limitations must be acknowledged. First, the simulation environment relies on specific assumptions such as perfect information (i.e. exact environmental conditions, asset status), and seamless collaboration between the centralized RL agent and the mission assets. In reality, relaying this information is challenging. The process riddled with delays and sometimes misinformation; particularly in missions with low visibility or competitive adversaries. Moreover, our approach assumed rapid detection of the threat within an hour and the action space was discretized into high-level tactical triplets (Selection, Tracking, and Suppression) rather than a continuous representation of possible operational actions. While this abstraction allows for efficient learning, it abstracts the agent's ability to execute specialized actions that might be possible to represent in a continuous control environment. Furthermore, we only experimented with a fixed set of homogenous assets for simplicity; however, the ideas illustrated here could be extended to larger sets of heterogenous assets. Hence the findings do not alter the main conclusion that RL can demonstrate efficacy in such high-fidelity operational decision making. Finally, while state-of-the-art RL applications often require extensive training periods and fine tuning, this study restricted the training to 3,000 simulations to establish a proof of concept despite high computational costs. Hence the performance gains noted in this work should be treated as a conservative estimate of the efficacy of the proposed method. 

\section{Conclusions and Future Work}

ME require rigorous analytical decision-making approaches for asset acquisition, fleet selection, and mission execution given a vast array of mission and SoS uncertainties. This renders traditional tradespace exploration and optimization methods computationally intractable, and calls for new perspectives. As a step towards achieving resilient mission outcomes, this study provides a proof of concept for integrating high-fidelity mission models with RL to enable intelligent and integrated mission coordination. By this method, policies for tactical decision-making can be trained over repetitive simulations, matching the states of the system with the appropriate tactical actions without explicitly modelling the uncertainty in the mission scenario. \\
\indent The problem tackled in this study was only concerned with the tactical management of a mission. The hope is to use this as a step to build work towards fleet selection and new system acquisition problems. This is because the ME landscape is composed of several hierarchies that are distributed across different time cycles, and in addition tactics management encompasses both the fleet selection, its sustainment, and design of new assets to be integrated into an existing fleet. Our long-term objective is to treat the methodology provided in this work as a basis for valuing ME approaches in over extended time periods. \\ 
\indent Future work in the near term address any of the limitations discussed earlier in this text such as imperfect information sharing and communication delays, to better reflect the challenges of real-world missions. Alternatively, this work could be expanded to account of the mission value of fleets with heterogenous assets. Another future direction is integrating explainable AI methodologies so that the trained policies are explainable and also provide more insight for both the practitioners and policy makers. In conclusion, we believe ME will remain a fruitful research venue on the intersection of SE and artificial intelligence, and we would like to encourage the SE community to expand into this area.

\bibliographystyle{unsrt}
\bibliography{FirefightingSoS}
\end{document}